\documentclass[onecolumn]{IEEEtran}
\usepackage{amsmath}
\usepackage{amssymb}
\usepackage{mathrsfs}
\usepackage{cite}
\usepackage{epsfig}
\usepackage{epsfig}
\usepackage{theorem}
\usepackage{graphics}
\usepackage{hyperref}
\usepackage{epsfig}

\newtheorem{thm}{Theorem}
\newtheorem{lem}{Lemma}

\newtheorem{defi}{Definition}

\newtheorem{rem}{Remark}

\begin{document}

\title{Secrecy Rate Region of the Broadcast Channel with an Eavesdropper }

\author{Ghadamali Bagherikaram, Abolfazl S. Motahari, Amir K. Khandani\\
Coding and Signal Transmission Laboratory,\\
 Department of
Electrical
and Computer Engineering,\\
 University of Waterloo, Waterloo, Ontario,
 N2L 3G1\\
 Emails: \{gbagheri,abolfazl,khandani\}@cst.uwaterloo.ca
}
 \maketitle
\footnote{Financial support provided by Nortel and the corresponding
matching funds by the Natural Sciences and Engineering Research
Council of Canada (NSERC), and Ontario Centres of Excellence (OCE)
are gratefully acknowledged.}
\begin{abstract}
In this paper, we consider a scenario where a source node wishes to
broadcast two confidential messages to two receivers, while a
wire-tapper also receives the transmitted signal. This model is
motivated by wireless communications, where individual secure
messages are broadcast over open media and can be received by any
illegitimate receiver. The secrecy level is measured by the
equivocation rate at the eavesdropper. We first study the general
(non-degraded) broadcast channel with an eavesdropper. We present an
inner bound on the secrecy capacity region for this model. This
inner bound is based on a combination of random binning, and the
Gelfand-Pinsker binning. We further study the situation in which the
channels are degraded. For the degraded broadcast channel with an
eavesdropper, we present the secrecy capacity region. Our achievable
coding scheme is based on Cover's superposition scheme and random
binning. We refer to this scheme as the Secret Superposition Scheme.
Our converse proof is based on a combination of the converse proof
of the conventional degraded broadcast channel and Csiszar Lemma. We
then assume that the channels are Additive White Gaussian Noise
(AWGN) and show that the Secret Superposition Scheme with Gaussian
codebook is optimal. The converse proof is based on Costa's entropy
power inequality. Finally, we use a broadcast strategy for the
slowly fading wire-tap channel when only the eavesdropper's channel
is fixed and known at the transmitter. We derive the optimum power
allocation for the coding layers, which maximizes the total average
rate.
\end{abstract}
\section{Introduction}
The notion of information theoretic secrecy in communication systems
was first introduced by Shannon in \cite{1}. The information
theoretic secrecy requires that the received signal of the
eavesdropper does not provide any information about the transmitted
messages. Shannon considered a pessimistic situation where both the
intended receiver and the eavesdropper have direct access to the
transmitted signal (which is called ciphertext). Under these
circumstances, he proved a negative result showing that perfect
secrecy can be achieved only when the entropy of the secret key is
greater than, or equal to, the entropy of the message. In modern
cryptography, all practical cryptosystems are based on Shannnon's
pessimistic assumption. Due to practical constraints, secret keys
are much shorter than messages; therefore, these practical
cryptosystems are theoretically susceptible to breaking by
attackers. The goal of designing such practical ciphers, however, is
to guarantee that no efficient algorithm exists for breaking them.

Wyner in \cite{2} showed that the above negative result is a
consequence of Shannon's restrictive assumption that the adversary
has access to precisely the same information as the legitimate
receiver. Wyner considered a scenario in which a wire-tapper
receives the transmitted signal over a degraded channel with respect
to the legitimate receiver's channel. He further assumed that the
wire-tapper has no computational limitations and knows the codebook
used by the transmitter. He measured the level of ignorance at the
eavesdropper by its equivocation and characterized the
capacity-equivocation region. Interestingly, a non-negative perfect
secrecy capacity is always achievable for this scenario.

The secrecy capacity for the Gaussian wire-tap channel is
characterized by Leung-Yan-Cheong in \cite{3}. Wyner's work is then
extended to the general (non-degraded) broadcast channel with
confidential messages by Csiszar and Korner \cite{4}. They
considered transmitting confidential information to the legitimate
receiver while transmitting common information to both the
legitimate receiver and the wire-tapper. They established a
capacity-equivocation region for this channel. The BCC has recently
been further studied in \cite{5,6,7}, where the source node
transmits a common message to both receivers, along with two
additional confidential messages, each aimed at one of the two
receivers. Here, the confidentiality of each message is measured
with respect to the other user, and there is no external
eavesdropper.

The fading wire-tap channel is investigated in \cite{8} where the
source-to destination channel and the source-to-eavesdropper channel
are corrupted by multiplicative fading gain coefficients, in
addition to additive white Gaussian noise. In this work, channels
are fast fading and the Channel State Information (CSI) of the
legitimate receiver is available at the transmitter. The perfect
secrecy capacity is derived for two different scenarios regarding
the availability of the eavesdropper's CSI. Moreover, the optimal
power control policy is obtained for the different scenarios. The
effect of the slowly fading channel on the secrecy capacity of a
conventional wire-tap channel is studied in \cite{9,10}. In these
works, it is assumed that the fading is quasi-static and the
transmitter does not know the fading gains. The outage probability,
which is the probability that the main channel is stronger than the
eavesdropper's channel, is defined in these works. In an outage
strategy, the transmission rate is fixed and the information is
detected when the instantaneous main channel is stronger than the
instantaneous eavesdropper's channel; otherwise, either nothing is
decoded at the legitimate receiver, or the information is leaked to
the eavesdropper. The term outage capacity refers to the maximum
achievable average rate. In \cite{10_1}, a broadcast strategy for
the slowly fading Gaussian point to point channel is introduced. In
this strategy, the transmitter uses a layered coding scheme and the
receiver is viewed as a continuum of ordered users.

In \cite{11}, the wire-tap channel is extended to the parallel
broadcast channels and also to the fading channels with multiple
receivers. In \cite{11}, the secrecy constraint is a perfect
equivocation for each of the messages, even if all the other
messages are revealed to the eavesdropper. The secrecy sum capacity
for a reverse broadcast channel is derived subject to this
restrictive assumption. The notion of the wire-tap channel is also
extended to multiple access channels \cite{12,13,14,15}, relay
channels \cite{16,17,18,19}, parallel channels \cite{20} and
Multiple-Input Multiple-Output channels
\cite{21,22,23,24,25,26,26_2}. Some other related works on the
communication of confidential messages can be found in
\cite{27,28,29,30,31}.

In this paper, we consider a scenario where a source node wishes to
broadcast two confidential messages to two receivers, while a
wire-tapper also receives the transmitted signal. This model is
motivated by wireless communications, where individual secure
messages are broadcast over shared media and can be received by any
illegitimate receiver. In fact, we simplify the restrictive
constraint imposed in \cite{11} and assume that the eavesdropper
does not have access to the other messages. We first study the
general broadcast channel with an eavesdropper. We present an
achievable rate region for this channel. Our achievable coding
scheme is based on a combination of random binning and the
Gelfand-Pinsker binning \cite{32}. This scheme matches the Marton's
inner bound \cite{33} on the broadcast channel without
confidentiality constraint. We further study the situation where the
channels are physically degraded and characterize the corresponding
secrecy capacity region. Our achievable coding scheme is based on
Cover's superposition coding \cite{34} and random binning. We refer
to this scheme as the Secret Superposition Coding. This capacity
region matches the capacity region of the degraded broadcast channel
without any security constraint. It also matches the secrecy
capacity of the wire-tap channel. We also characterize the secrecy
capacity region when the channels are additive white Gaussian noise.
We show that the secret superposition of Gaussian codebooks is the
optimal choice. Based on the rate characterization of the secure
broadcast channel, we then use broadcast strategy for the slow
fading wire-tap channel when only the eavesdropper's channel is
fixed and known at the transmitter. In broadcast strategy, a source
node sends secure layers of coding and the receiver is viewed as a
continuum of ordered users. We derive optimum power allocation for
the layers which maximizes the total average rate.

In \cite{35}, we published a conference version of this work where
the achievable rate region of the general broadcast channel with an
eavesdropper and the secrecy capacity region of the degraded one
were addressed. However, we later became aware that reference
\cite{36,37} had considered a similar model as used in this paper
and had independently characterized the secrecy capacity region of
the broadcast channel (when the channels are degraded). They also
generalized their results to the parallel degraded broadcast channel
with an eavesdropper. Independently and parallel to our work,
reference \cite{38} considered the Gaussian broadcast channel with
an eavesdropper and characterized its capacity region. Authors of
\cite{38} provided two methods for their converse proof. The first
one uses the alternative representation of the mutual information as
an integration of the minimum-mean-square-error (MMSE), as well as
the properties of the MMSE. The second one uses the relationship
between the differential entropy and the Fisher information via the
de Bruin identity, along with the properties of the Fisher
information. In this work, however, we use Costa's entropy power
inequality to provide the converse proof.

The rest of the paper is organized as follows: in section II we
introduce the system model. In section III we provide an inner bound
on the secrecy capacity region when the channels are not degraded.
In section IV we specialize our channel to the degraded ones and
establish the secrecy capacity region. In section V we derive the
secrecy capacity region when the channels are AWGN. Based on the
secrecy capacity region of the AWGN channel, in section VI we use a
broadcast strategy for the slow fading wire-tap channel when the
transmitter only knows the eavesdropper's channel. Finally, section
VII concludes the paper.

\section{Preliminaries}
In this paper, random variables are denoted by capital letters
(e.g. $X$) and their realizations are denoted by corresponding lower
case letters (e.g. $x$). The finite alphabet of a random variable is
denoted by a script letter (e.g. $\mathcal{X}$) and its probability
distribution is denoted by $P(x)$. The
vectors will be written as
$x^{n}=(x_{1},x_{2},...,x_{n})$, where subscripted letters denote
the components and superscripted letters denote the vector. Bold capital letters represent matrices (e.g. $\mathbf{A}$). The
notation $x^{i-1}$ denotes the vector $(x_{1},x_{2},...,x_{i-1})$
and the notation $\widetilde{x}^{i}$ denotes the vector
$(x_{i},x_{i+1},...,x_{n})$. A similar notation will be used for
random variables and random vectors.

\begin{figure}
\centerline{\includegraphics[scale=.6]{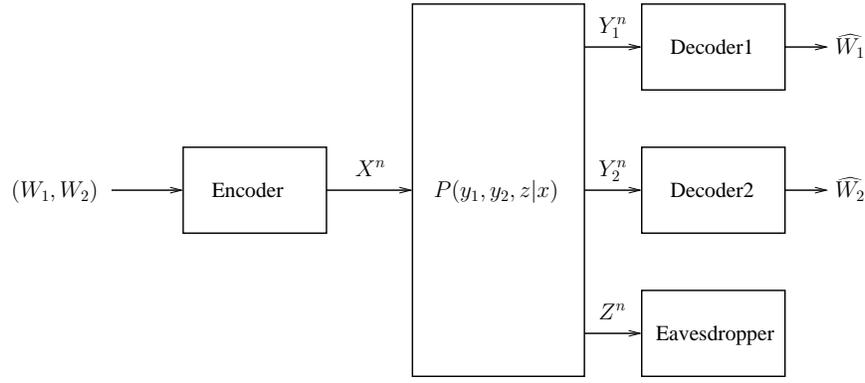}} \caption{Broadcast
Channel with an Eavesdropper} \label{f1}
\end{figure}

Consider a Broadcast Channel with an eavesdropper (BCE) as depicted
in Fig. \ref{f1}.
 In this confidential setting, the transmitter wishes to send two independent messages
$(W_{1},W_{2})$ to the respective receivers in $n$ uses of the
channel and prevent the eavesdropper from having any information
about the messages. A discrete memoryless broadcast channel with an
eavesdropper is represented by
$(\mathcal{X},P,\mathcal{Y}_{1},\mathcal{Y}_{2},\mathcal{Z})$ where,
$\mathcal{X}$ is the finite input alphabet set, $\mathcal{Y}_{1}$,
$\mathcal{Y}_{2}$ and $\mathcal{Z}$ are three finite output alphabet
sets, and $P$ is the channel transition probability
$P(y_{1},y_{2},z|x)$. The input of the channel is
$x^{n}\in\mathcal{X}^{n}$ and the outputs are
$y_{1}^{n}\in\mathcal{Y}_{1}^{n}$, $y_{2}^{n}\in
\mathcal{Y}_{2}^{n},$ and $z^{n}\in \mathcal{Z}^{n}$ for Receiver
$1$, Receiver $2$, and the eavesdropper, respectively. The channel
is discrete memoryless in the sense that
\begin{equation}
P(y_{1}^{n},y_{2}^{n},z^{n}|x^{n})=\prod_{i=1}^{n}P(y_{1,i},y_{2,i},z_{i}|x_{i}).
\end{equation}
A $((2^{nR_{1}},2^{nR_{2}}),n)$ code for a broadcast channel with an
eavesdropper consists of a stochastic encoder
\begin{equation}
f:(\{1,2,...,2^{nR_{1}}\}\times\{1,2,...,2^{nR_{2}}\})\rightarrow
\mathcal{X}^{n},
\end{equation}
and two decoders,
\begin{equation}
g_{1}:\mathcal{Y}_{1}^{n}\rightarrow \{1,2,...,2^{nR_{1}}\}
\end{equation}
and
\begin{equation}
g_{2}:\mathcal{Y}_{2}^{n}\rightarrow \{1,2,...,2^{nR_{2}}\}.
\end{equation}
The average probability of error is defined as the probability that
the decoded messages are not equal to the transmitted messages; that
is,
\begin{equation}
P_{e}^{(n)}=P(g_{1}(Y_{1}^{n})\neq W_{1}\cup g_{2}(Y_{2}^{n})\neq
W_{2}).
\end{equation}

The knowledge that the eavesdropper can extract about $W_{1}$ and
$W_{2}$ from its received signal $Z^{n}$ is measured by
\begin{IEEEeqnarray}{lr}
I(Z^{n},W_{1})=H(W_{1})-H(W_{1}|Z^{n}),\\
I(Z^{n},W_{2})=H(W_{2})-H(W_{2}|Z^{n}),
\end{IEEEeqnarray}
and
\begin{equation}
I(Z^{n},(W_{1},W_{2}))=H(W_{1},W_{2})-H(W_{1},W_{2}|Z^{n}).
\end{equation}
Perfect secrecy revolves around the idea that the eavesdropper
should not obtain any information about the transmitted messages.
Perfect secrecy thus requires that
\begin{IEEEeqnarray}{lr}\nonumber
I(Z^{n},W_{1})=0\Leftrightarrow H(W_{1})=H(W_{1}|Z^{n}),\\
\nonumber I(Z^{n},W_{2})=0\Leftrightarrow H(W_{2})=H(W_{2}|Z^{n}),
\end{IEEEeqnarray}
and
\begin{equation}\nonumber
I(Z^{n},(W_{1},W_{2}))=0\Leftrightarrow
H(W_{1},W_{2})=H(W_{1},W_{2}|Z^{n}).
\end{equation}
where $n\rightarrow\infty$. The secrecy levels of confidential
messages $W_{1}$ and $W_{2}$ are measured at the eavesdropper in
terms of equivocation rates which are defined as follows:
\begin{defi}
The equivocation rates $R_{e1}$, $R_{e2}$ and $R_{e12}$
 for the
broadcast channel with an eavesdropper are:
\begin{IEEEeqnarray}{lr}\nonumber
R_{e1}=\frac{1}{n}H(W_{1}|Z^{n}),\\
\nonumber R_{e2}=\frac{1}{n}H(W_{2}|Z^{n}), \\ \nonumber
R_{e12}=\frac{1}{n}H(W_{1},W_{2}|Z^{n}).
\end{IEEEeqnarray}
\end{defi}
The perfect secrecy rates $R_{1}$ and $R_{2}$ are the amount of
information that can be sent to the legitimate receivers in a
reliable and confidential manner.
\begin{defi}
A secrecy rate pair $(R_{1},R_{2})$ is said to be achievable if for
any $\epsilon>0,\epsilon_{1}>0,\epsilon_{2}>0,\epsilon_{3}>0$, there
exists a sequence of $((2^{nR_{1}},2^{nR_{2}}),n)$ codes, such that
for sufficiently large $n$, we have:
\begin{IEEEeqnarray}{rl}
\label{l0}P_{e}^{(n)}&\leq \epsilon,\\
\label{l1}
 R_{e1}&\geq R_{1}-\epsilon_{1},\\
\label{l2} R_{e2}&\geq R_{2}-\epsilon_{2},\\
\label{l3}
 R_{e12}&\geq R_{1}+R_{2}-\epsilon_{3}.
\end{IEEEeqnarray}
\end{defi}
In the above definition, the first condition concerns the
reliability, while the other conditions guarantee perfect secrecy
for each individual message and the combination of the two messages,
respectively. Since the messages are independent of each other, the
conditions of (\ref{l1}) and (\ref{l3}) or (\ref{l2}) and (\ref{l3})
are sufficient to provide perfect secrecy.
 The capacity region is defined as follows.
\begin{defi}
The capacity region of the broadcast channel with an eavesdropper is
the closure of the set of all achievable rate pairs $(R_{1},R_{2})$.
\end{defi}

\section{Achievable Rates for General BCE}
In this section, we consider the general broadcast channel with an
eavesdropper and present an achievable rate region. Our achievable
coding scheme is based on a combination of the random binning,
superposition coding, rate splitting, and Gelfand-Pinsker binning
schemes \cite{32}. Our binning approach is supplemented with
superposition coding to accommodate the common message. We call this
scheme the Secret Superposition Scheme. An additional binning is
introduced for the confidentiality of private messages. We note that
these double binning techniques have been used by various authors
for secret communication (see e.g. \cite{5,7}). The following
theorem illustrates the achievable rate region for this channel.
\begin{thm}\label{th1}
Let $\mathbb{R}_{I}$ denote the union of all non-negative rate pairs
$(R_{0},R_{1},R_{2})$ satisfying
 \begin{IEEEeqnarray}{rl}\nonumber
R_{0}&\leq \min\{I(U;Y_{1}),I(U;Y_{2})\}-I(U;Z),\\ \nonumber
R_{1}+R_{0}&\leq
I(V_{1};Y_{1}|U)-I(V_{1};Z|U)+\min\{I(U;Y_{1}),I(U;Y_{2})\}-I(U;Z),
\\ \nonumber R_{2}+R_{0}&\leq I(V_{2};Y_{2}|U)-I(V_{2};Z|U)+\min\{I(U;Y_{1}),I(U;Y_{2})\}-I(U;Z), \\
\nonumber R_{1}+R_{2}+R_{0}&\leq
    I(V_{1};Y_{1}|U)+I(V_{2};Y_{2}|U)-I(V_{1},V_{2};Z|U)-I(V_{1};V_{2}|U)+\min\{I(U;Y_{1}),I(U;Y_{2})\}-I(U;Z),
\end{IEEEeqnarray}
over all joint distributions
$P(u)P(v_{1},v_{2}|u)P(x|v_{1},v_{2})P(y_{1},y_{2},z|x)$. Any rate
pair $(R_{0},R_{1},R_{2})\in \mathbb{R}_{I}$ is then achievable for
the broadcast channel with an eavesdropper and with common
information.
\end{thm}
Please see Appendix \ref{app1} for the proof.
\begin{rem}
If we remove the secrecy constraints by removing the eavesdropper,
the above rate region becomes Marton's achievable region with common
information for the general broadcast channel.
\end{rem}
\begin{rem}
If we remove one of the users, e.g. user $2$ and the common message,
then we get Csiszar and Korner's secrecy capacity for the other
user.
\end{rem}

\section{The Capacity Region of the Degraded BCE}
In this section, we consider the degraded broadcast channel with an
eavesdropper and establish its secrecy capacity region.
\begin{defi}
A broadcast channel with an eavesdropper is said to be physically
degraded, if $X\rightarrow Y_{1} \rightarrow Y_{2}\rightarrow Z$
forms a Markov chain. In other words, we have
\begin{equation}
P(y_{1},y_{2},z|x)=P(y_{1}|x)P(y_{2}|y_{1})P(z|y_{2}).\nonumber
\end{equation}
\end{defi}
\begin{defi}
A broadcast channel with an eavesdropper is said to be
stochastically degraded if its conditional marginal distributions
are the same as that of a physically degraded broadcast channel,
i.e., if there exist two distributions $P^{'}(y_{2}|y_{1})$  and
$P^{'}(z|y_{2})$, such that

\begin{IEEEeqnarray}{rl}
\nonumber P(y_{2}|x)&=\sum_{y_{1}}P(y_{1}|x)P^{'}(y_{2}|y_{1}),\\
\nonumber P(z|x)&=\sum_{y_{2}}P(y_{2}|x)P^{'}(z|y_{2}).
\end{IEEEeqnarray}
\end{defi}
\begin{lem}
The secrecy capacity region of a broadcast channel with an
eavesdropper depends only on the conditional marginal distributions
$P(y_{1}|x)$, $P(y_{2}|x)$ and $P(z|x)$.
\end{lem}
\begin{proof}
It suffices to show that the error probability of  $P_{e}^{(n)}$ and
the equivocations of $H(W_{1}|Z^{n})$, $H(W_{2}|Z^{n})$ and
$H(W_{1},W_{2}|Z^{n})$ are only  functions of the marginal
distributions when we use the same codebook and encoding schemes.
Note that
\begin{equation}\nonumber
\max\{P_{e,1}^{(n)},P_{e,2}^{(n)}\}\leq P_{e}^{(n)}\leq
P_{e,1}^{(n)}+P_{e,2}^{(n)}.
\end{equation}
Hence, $P_{e}^{(n)}$ is small if, and only if, both $P_{e,1}^{(n)}$
and $P_{e,2}^{(n)}$ are small. On the other hand, for a given
codebook and encoding scheme, the decoding error probabilities
$P_{e,1}^{(n)}$ and $P_{e,2}^{(n)}$ and the equivocation rates
depend only on marginal channel probability densities $P_{Y_{1}|X}$,
$P_{Y_{2}|X}$ and $P_{Z|X}$. Thus, the same code and encoding scheme
gives the same $P_{e}^{(n)}$ and equivocation rates.
\end{proof}

In the following theorem, we fully characterize the capacity region
of the physically degraded broadcast channel with an eavesdropper.
\begin{thm} \label{th2}
The capacity region for transmitting independent secret information
over the degraded broadcast channel is the convex hull of the
closure of all $(R_{1},R_{2})$ satisfying
\begin{IEEEeqnarray}{rl}\label{l7}
    R_{1}&\leq I(X;Y_{1}|U)-I(X;Z|U), \\ \label{l7_1}
    R_{2}&\leq I(U;Y_{2})-I(U;Z),
\end{IEEEeqnarray}
for some joint distribution $P(u)P(x|u)P(y_{1}, y_{2},z|x)$.
\end{thm}
Please refer to Appendix \ref{app2} for the proof.
\begin{rem}
If we remove the secrecy constraints by removing the eavesdropper, then the above theorem becomes the capacity
region of the degraded broadcast channel.
\end{rem}
The coding scheme is based on Cover's superposition coding and
random binning. We refer to this scheme as the Secure Superposition
Coding scheme. The available resources at the encoder are used for
two purposes: to confuse the eavesdropper so that perfect secrecy
can be achieved for both layers, and to transmit the messages into
the main channels. To satisfy confidentiality, the randomization
used in the first layer is fully exploited in the second layer. This
makes an increase of $I(U;Z)$ in the bound of $R_{1}$.

\begin{rem}
As Lemma \ref{ll1} bounds the secrecy rates for the general
broadcast channel with an eavesdropper then, Theorem \ref{th2} is
true when only the legitimate receivers are degraded.
\end{rem}

\section{Capacity Region of Gaussian BCE}
In this section, we consider the Gaussian Broadcast Channel with an
Eavesdropper (G-BCE). Note that optimizing (\ref{l7}) and
(\ref{l7_1}) for AWGN channels involves solving a nonconvex
functional. Usually nontrivial techniques and strong inequalities
are used to solve the optimization problems of this type. In
\cite{3}, Leung-Yan-Cheong successfully evaluated the capacity
expression of the wire-tap channel by using the entropy power
inequality \cite{38_1,38_2}. Alternatively, it can also be evaluated
using a classical result from the Estimation Theory and the
relationship between mutual information and minimum mean-squared
error estimation. On the other hand, the entropy power inequality is
sufficient to establish the converse proof of a Gaussian broadcast
channel without secrecy constraint. Unfortunately, the traditional
entropy power inequality does not extend to the secure multi-user
case. Here, by using Costa's version of the entropy power
inequality, we show that secret superposition coding with Gaussian
codebook is optimal.

\begin{figure}
\centerline{\includegraphics[scale=.6]{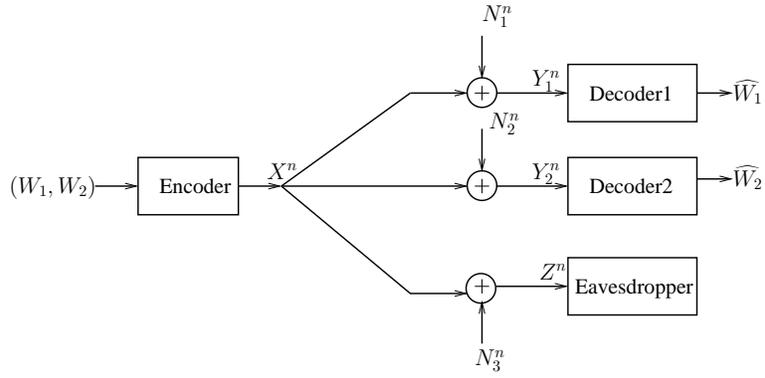}} \caption{Gaussian
Broadcast Channel with an Eavesdropper(G-BCE)} \label{f4_1}
\end{figure}

\begin{figure}
\centerline{\includegraphics[scale=.6]{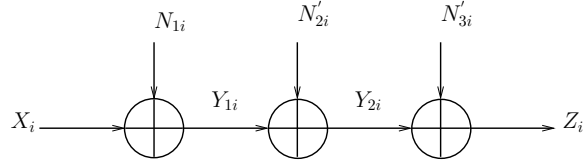}}
\caption{Equivalent Channels for the G-BCE}  \label{f4}
\end{figure}

Figure~\ref{f4_1} shows the channel model. At time $i$ the received
signals are $Y_{1i}=X_{i}+N_{1i}$, $Y_{2i}=X_{i}+N_{2i}$ and
$Z_{i}=X_{i}+N_{3i}$, where $N_{ji}$ is a Gaussian random variable
with zero mean and $Var(N_{ji})=\sigma_{j}^{2}$ for $j=1,2,3$. Here
$\sigma_{1}^{2}\leq \sigma_{2}^{2}\leq \sigma_{3}^{2}$. Assume that
the transmitted power is limited to $E[X^{2}]\leq P$.
Since the channels are degraded, the received signals can
alternatively be written as $Y_{1i}=X_{i}+N_{1i}$,
$Y_{2i}=Y_{1i}+N_{2i}^{'}$ and $Z_{i}=Y_{2i}+N_{3i}^{'}$, where
$N_{1i}$'s are i.i.d $\mathcal{N}(0,\sigma_{1}^{2})$, $N_{2i}^{'}$'s
are i.i.d $\mathcal{N}(0,\sigma_{2}^{2}-\sigma_{1}^{2})$, and
$N_{3i}^{'}$'s are i.i.d
$\mathcal{N}(0,\sigma_{3}^{2}-\sigma_{2}^{2})$. Fig.~\ref{f4} shows
the equivalent channels for the G-BCE. The following theorem
illustrates the secrecy capacity region of G-BCE.
\begin{thm}\label{th3}
The secrecy capacity region of the G-BCE is given by the set of
rates pairs $(R_{1},R_{2})$ satisfying
\begin{IEEEeqnarray}{rl}\label{g1}
    R_{1}&\leq C\left(\frac{\alpha P}{\sigma_{1}^{2}}\right)-C\left(\frac{\alpha P}{\sigma_{3}^{2}}\right),
    \\ \label{g2}
    R_{2}&\leq C\left(\frac{(1-\alpha) P}{\alpha P +\sigma_{2}^{2}}\right)-C\left(\frac{(1-\alpha) P}{\alpha P +\sigma_{3}^{2}}\right).
\end{IEEEeqnarray}
for some $\alpha \in [0,1]$.
\end{thm}
Please see Appendix \ref{app3} for the proof.

\begin{figure}
\centerline{\includegraphics[scale=.5]{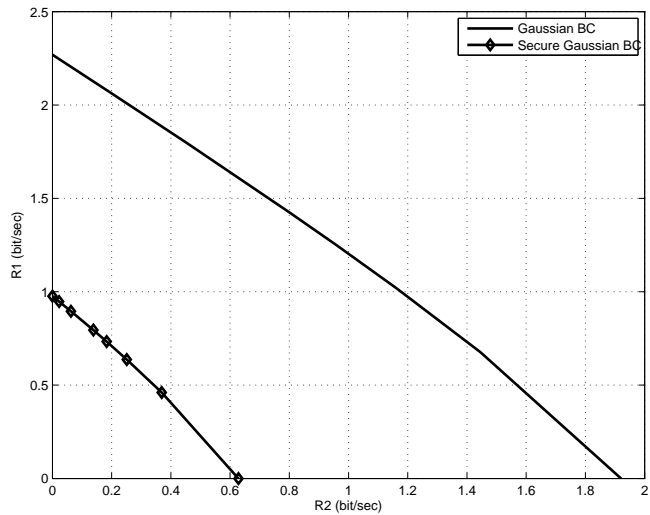}}
\caption{Secret/Non-Secret Capacity Region of a Degraded Broadcast
Channel } \label{f5}
\end{figure}

Figure \ref{f5} shows the capacity region of a degraded Gaussian
broadcast channel with and without secrecy constraint. In this
figure $P=20$, $N_{1}=0.9$, $N_{2}=1.5$ and $N_{3}=4$.

\section{A Multilevel Coding Approach to the Slowly Fading Wire-Tap Channel}
In this section, we use the secure degraded broadcast channel from
the previous section to develop a new broadcast strategy for a slow
fading wire-tap channel. This strategy aims to maximize the average
achievable rate where the main channel state information is not
available at the transmitter. By assuming that there is an infinite
number of ordered receivers which correspond to different channel
realizations, we propose a secret multilevel coding scheme that
maximizes the underlying objective function. First, some
preliminaries and definitions are given, and then the proposed
multilevel coding scheme is described. Here, we follow the steps of
the broadcast strategy for the slowly fading point-to-point channel
of \cite{10_1}. This method is used in several other papers; see,
e.g, \cite{10_2,10_3,10_4}.
\subsection{Channel Model}
\begin{figure}
\centerline{\includegraphics[scale=.6]{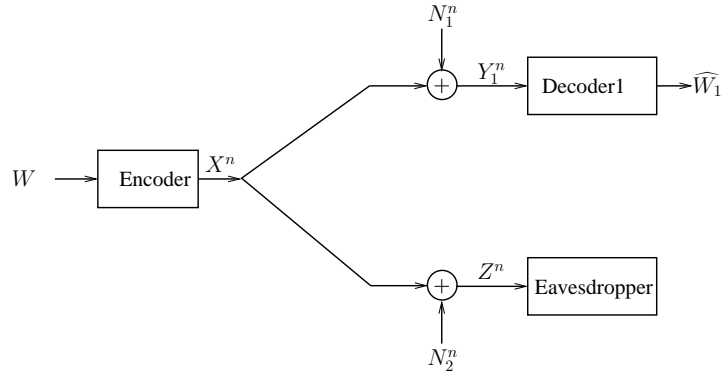}} \caption{Gaussian
Wire-tap Channel } \label{f7}
\end{figure}

Consider a wire-tap channel as depicted in Fig.\ref{f7}.
The transmitter wishes to communicate with the destination in the
presence of an eavesdropper. At time $i$, the signal received
by the destination and the eavesdropper are given as follows
\begin{IEEEeqnarray}{lr}\label{eq2}
Y_{i}=h_{M}X_{i}+N_{1i}\\
\nonumber Z_{i}=h_{E}X_{i}+N_{2i}
\end{IEEEeqnarray}
where  $X_{i}$ is the transmitted symbol and $h_{M}$, $h_{E}$ are
the fading coefficients from the source to the legitimate receiver
and to the eavesdropper, respectively. The fading power gains of the
main and eavesdropper channels are given by $s=|h_{M}|^{2}$ and
$s^{'}=|h_{E}|^{2}$, respectively. $N_{1i}$, $N_{2i}$ are the
additive noise samples, which are Gaussian i.i.d with zero mean and
unit variance. We assume that the channels are slowly fading, and
also assume that the transmitter knows only  channel state
information of the eavesdropper channel. A motivation for this
assumption is that when both channels are unknown at the
transmitter, we assume that $s^{'}=|h_{E}|^{2}$ denotes the
best-case eavesdropper channel gain. For each realization of $h_{M}$
there is an achievable rate. Since the transmitter has no
information about the main channel and the channels are slowly
fading, then the system is non-ergodic. Here, we are interested in
the average rate for various independent transmission blocks. The
average shall be calculated over the distribution of $h_{M}$.
\subsection{The Secret Multilevel Coding Approach}
\begin{figure}
\centerline{\includegraphics[scale=.6]{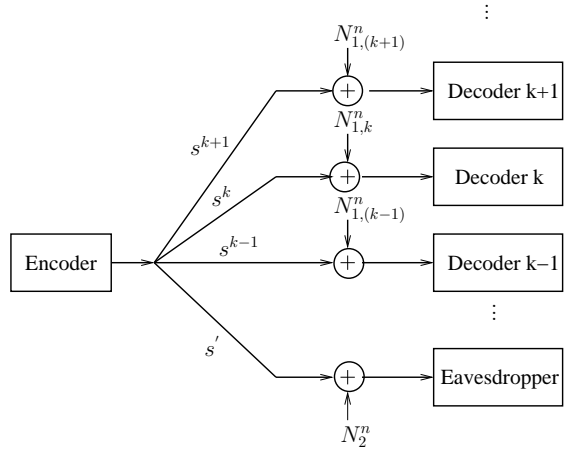}} \caption{ Equivalent Broadcast Channel Model. } \label{f6}
\end{figure}

An equivalent broadcast channel for our channel is depicted in
Fig.~\ref{f6}.
wherein the transmitter sends an infinite number of secure layers of
coded information. The receiver is equivalent to a continuum of
ordered users. For each channel realization $h_{M}^{k}$ with the
fading power gain $s^{k}$, the information rate is $R(s^{k},s^{'})$.
We drop the superscript $k$, and the realization of the fading power
random variable $S$  is denoted by $s$. Therefore, the transmitter
views the main channel as a secure degraded Gaussian broadcast
channel with an infinite number of receivers. The result of the
previous section for the two receivers can easily be extended to an
arbitrary number of users. According to theorem \ref{th3}, the
incremental differential secure rate is then given by
\begin{eqnarray}
dR(s,s^{'})&=& \left[\frac{1}{2}
\log\left(1+\frac{s\rho(s)ds}{1+sI(s)}\right)-\frac{1}{2}\log\left(1+\frac{s^{'}\rho(s)ds}{1+s^{'}I(s)}\right)\right]^{+},
\end{eqnarray}
where $\rho(s)ds$ is the transmit power of a layer parameterized by
$s$, intended for receiver $s$. As $\log(1+x)\approx x$ for $x\leq
1$ then the $\log$ function may be discarded. The function $I(s)$
represents the interference noise of the receivers indexed by $u>s$
which cannot be canceled at receiver $s$. The interference at
receiver $s$ is therefore given by
\begin{equation}
I(s)=\int_{s}^{\infty}\rho(u)d(u).
\end{equation}
The total transmitted power is the summation of the power assigned
to the layers
\begin{equation}\label{pc}
P=I(0)=\int_{0}^{\infty}\rho(u)d(u).
\end{equation}
The total achievable rate for a fading realization $s$ is an
integration of the incremental rates over all receivers, which can
successfully decode the respective layer
\begin{IEEEeqnarray}{rl}
R(s,s^{'})=\frac{1}{2}\int_{0}^{s}\left[\frac{u\rho(u)du}{1+uI(u)}-\frac{s^{'}\rho(u)du}{1+s^{'}I(u)}\right]^{+}.
\end{IEEEeqnarray}
Our goal is to maximize the total average rate over all fading
realizations with respect to the power distribution $\rho(s)$ (or
equivalently, with respect to $I(u)$, $u\geq 0$) under the power
constraint  of \ref{pc}. The optimization problem may be written as
\begin{IEEEeqnarray}{rl}
R_{\max}&=\max_{I(u)}\int_{0}^{\infty}R(u,s^{'})f(u)du,\\ \nonumber &s.t\\
\nonumber P&=I(0)=\int_{0}^{\infty}\rho(u)d(u),
\end{IEEEeqnarray}
where $f(u)$ is the probability distribution function (pdf) of the
power gain $S$. Noting that the cumulative distribution function
(cdf) is $F(u)=\int_{0}^{u}f(a)da$, the optimization problem may be
written as
\begin{IEEEeqnarray}{rl}\label{fu}
R_{\max}&=\frac{1}{2}\max_{I(u)}\int_{0}^{\infty}(1-F(u))G(u)du,\\ \nonumber &s.t\\
\nonumber P&=I(0)=\int_{0}^{\infty}\rho(u)d(u),
\end{IEEEeqnarray}
where
$G(u)=\left[\frac{u}{1+uI(u)}-\frac{s^{'}}{1+s^{'}I(u)}\right]^{+}\rho(u)$.
Note that $\rho(u)=-I^{'}(u)$. Therefore, the functional in
(\ref{fu}) may be written as
\begin{IEEEeqnarray}{rl}\nonumber
&J(x,I(x),I^{'}(x))=\\
&-(1-F(x))\left[\frac{x}{1+xI(x)}-\frac{s^{'}}{1+s^{'}I(x)}\right]^{+}I^{'}(x).
\end{IEEEeqnarray}
The necessary condition for the maximization of an integral of $J$
over $x$ is
\begin{equation}
J_{I}-\frac{d}{dx}J_{I^{'}}=0,
\end{equation}
where $J_{I}$ means the derivation of function $J$ with respect to
$I$, and similarly $J_{I^{'}}$ is the derivation of $J$ with respect
to $I^{'}$. After some manipulations, the optimum $I(x)$ is given by
\begin{IEEEeqnarray}{rl}\nonumber
I(x)=\left\{
              \begin{array}{ll}
                \frac{1-F(x)-(x-s^{'})f(x)}{s^{'}(1-F(x))+x(x-s^{'})f(x)}, & \max\{s^{'},x_{0}\}\leq x \leq x_{1}; \\
                0, & \hbox{otherwise,}
              \end{array}
            \right.
\end{IEEEeqnarray}
where $x_{0}$ is determined by $I(x_{0})=P$, and $x_{1}$ by
$I(x_{1})=0$.

As a special case, consider the Rayleigh flat fading channel. The
random variable $S$ is exponentially distributed with
\begin{equation}
f(s)=e^{-s},~~~~~ F(s)=1-e^{-s},~~~~~ s\geq 0.
\end{equation}
Substituting  $f(s)$ and $F(s)$ into the optimum $I(s)$ and taking
the derivative with respect to the fading power $s$ yields the
following optimum transmitter power policy
\begin{IEEEeqnarray}{rl}\nonumber
\rho(s)=-\frac{d}{ds}I(s)=\left\{
              \begin{array}{ll}
                \frac{-s^{2}+2(s^{'}+1)s-s^{'2}}{(s^{2}-s^{'}s+s^{'})^{2}}, & \max\{s^{'},s_{0}\}\leq s \leq s_{1}; \\
                0, & \hbox{otherwise,}
              \end{array}
            \right.
\end{IEEEeqnarray}
where $s_{0}$ is the solution of the equation $I(s_{0})=P$, which is
\begin{equation}\nonumber
s_{0}=\frac{-1+Ps^{'}+\sqrt{P^{2}s^{'2}+2P(1-2P)s^{'}+4P+1}}{2P},
\end{equation}
and $s_{1}$ is determined by $I(s_{1})=0$, which is
\begin{equation}\nonumber
s_{1}=1+s^{'}.
\end{equation}
\section{Conclusion}
A generalization of the wire-tap channel in the case of two
receivers and one eavesdropper was considered. We established an
inner bound for the general (non-degraded) case. This bound matches
Marton's bound on broadcast channels without security constraint.
Furthermore, we considered the scenario in which the channels are
degraded. We established the perfect secrecy capacity region for
this case. The achievability coding scheme is a secret superposition
scheme where randomization in the first layer helps the secrecy of
the second layer. The converse proof combines the converse proof for
the degraded broadcast channel without security constraint, and the
perfect secrecy constraint. We proved that the secret superposition
scheme with the Gaussian codebook is optimal in AWGN-BCE. The
converse proof is based on Costa's entropy power inequality and
Csiszar lemma. Based on the rate characterization of the AWGN-BCE,
the broadcast strategy for the slowly fading wire-tap channel were
used. In this strategy, the transmitter only knows the
eavesdropper's channel and the source node sends secure layered
coding. The receiver is viewed as a continuum of ordered users. We
derived the optimum power allocation for the layers, which maximizes
the total average rate. \appendices
\section{Proof of Theorem \ref{th1}}\label{app1}
We split the private message $W_{1}\in\{1,2,...,2^{nR_{1}}\}$ into
$W_{11}\in\{1,2,...,2^{nR_{11}}\}$ and
$W_{10}\in\{1,2,...,2^{nR_{10}}\}$, and
$W_{2}\in\{1,2,...,2^{nR_{2}}\}$ into
$W_{22}\in\{1,2,...,2^{nR_{22}}\}$ and
$W_{20}\in\{1,2,...,2^{nR_{20}}\}$, respectively. $W_{11}$ and
$W_{22}$ are only to be decoded by the intended receivers, while
$W_{10}$ and $W_{20}$ are to be decoded by both receivers. Now, we
combine $(W_{10},W_{20},W_{0})$ into a single auxiliary variable
$U$. The messages $W_{11}$ and $W_{22}$ are represented by auxiliary
variables $V_{1}$ and $V_{2}$, respectively. Here,
$R_{10}+R_{11}=R_{1}$ and $R_{20}+R_{22}=R_{2}$.

\begin{figure}
\centerline{\includegraphics[scale=.6]{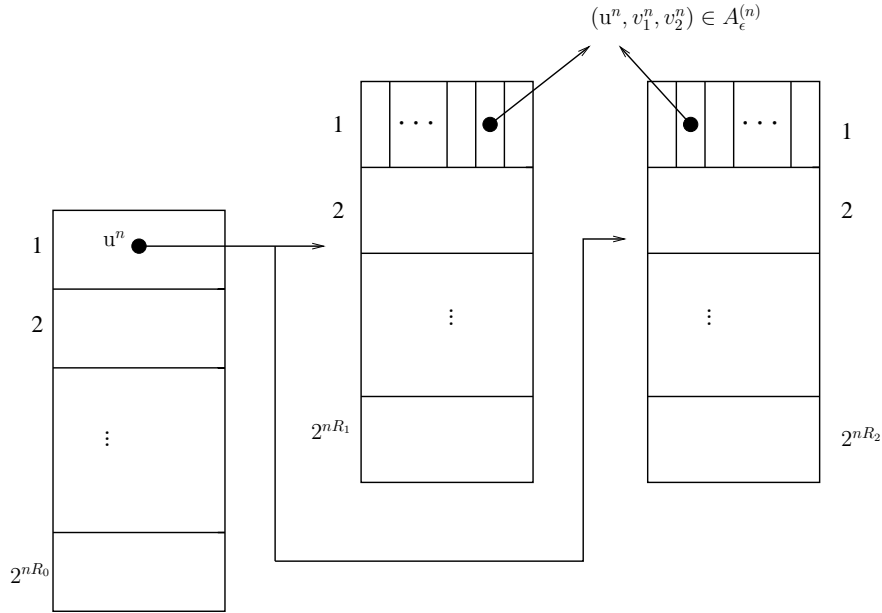}} \caption{The
Stochastic Encoder} \label{f2}
\end{figure}

1) \textit{Codebook Generation}:
 The structure of the encoder is
depicted in Fig.\ref{f2}.
Fix $P(u)$, $P(v_{1}|u)$, $P(v_{2}|u)$ and  $P(x|v_{1},v_{2})$. The
stochastic encoding is as follows. Define
\begin{IEEEeqnarray}{rl}\nonumber
L_{11}&=I(V_{1};Y_{1}|U)-I(V_{1};Z,V_{2}|U), \\ \nonumber
L_{12}&=I(V_{1};Z|V_{2},U),\\ \nonumber L_{21}&=I(V_{2};Z|V_{1},U)\\
\nonumber L_{22}&=I(V_{2};Y_{2}|U)-I(V_{2};Z,V_{1}|U),\\ \nonumber
L_{3}&=I(V_{1};V_{2}|U)-\epsilon,
\end{IEEEeqnarray}
Note that,
\begin{IEEEeqnarray}{rl}\nonumber
L_{11}+L_{12}+L_{3}&=I(V_{1};Y_{1}|U)-\epsilon, \\ \nonumber
L_{22}+L_{21}+L_{3}&=I(V_{2};Y_{2}|U)-\epsilon,
\end{IEEEeqnarray}
We first prove the case where
\begin{IEEEeqnarray}{rl}\label{ach}
R_{11}&\geq L_{11}\geq 0, \\
R_{22}&\geq L_{22}\geq 0.
\end{IEEEeqnarray}
Generate $2^{n(R_{10}+R_{20}+R_{0})}$ independent and identically
distributed (i.i.d) sequences $u^{n}(k)$ with
$k\in\{1,2,...,2^{R_{10}+R_{20}+R_{0}}\}$, according to the
distribution $P(u^{n})=\prod_{i=1}^{n}P(u_{i})$. For each codeword
$u^{n}(k)$, generate $2^{L_{11}+L_{12}+L_{3}}$ i.i.d codewords
$v_{1}^{n}(i,i^{'},i^{''})$, with $i\in\{1,2,...,2^{nL_{11}}\}$,
$i^{'}\in\{1,2,...,2^{nL_{12}}\}$ and
$i^{''}\in\{1,2,...,2^{nL_{3}}\}$, according to
$P(v_{1}^{n}|u^{n})=\prod_{i=1}^{n}P(v_{1i}|u_{i})$. The indexing
presents an alternative interpretation of binning. Randomly
distribute these sequences of $v_{1}^{n}$into $2^{nL_{11}}$ bins
indexed by $i$, for the codewords in each bin, randomly distribute
them into $2^{nL_{12}}$ sub-bins indexed by $i^{'}$; thus $i^{''}$
is the index for the codeword in each sub-bin. Similarly, for each
codeword $u^{n}$, generate $2^{L_{21}+L_{22}+L_{3}}$ i.i.d codewords
$v_{2}^{n}(j,j^{'},j^{''})$ according to
$P(v_{2}^{n}|u^{n})=\prod_{i=1}^{n}P(v_{2i}|u_{i})$, where
$j\in\{1,2,...,2^{nL_{21}}\}$, $j^{'}\in\{1,2,...,2^{nL_{22}}\}$ and
$j^{''}\in\{1,2,...,2^{nL_{3}}\}$.

2) \textit{Encoding}: To send messages $(w_{10},w_{20},w_{0})$, we
calculate the corresponding message index $k$ and choose the
corresponding codeword $u^{n}(k)$. Given this $u^{n}(k)$, there
exists $2^{n(L_{11}+L_{12}+L_{3})}$ codewords of
$v_{1}^{n}(i,i^{'},i^{''})$ to choose from for representing message
$w_{11}$. Evenly map $2^{nR_{11}}$ messages $w_{11}$ to
$2^{nL_{11}}$ bins, then, given (\ref{ach}), each bin corresponds to
at least one message $w_{11}$. Thus, given $w_{11}$, the bin index
$i$ can be decided.

\begin{enumerate}
  \item If $R_{11}\leq L_{11}+L_{12}$, each bin corresponds to
  $2^{n(R_{11}-L_{11})}$ messages $w_{11}$. Evenly place the
  $2^{nL_{12}}$ sub-bins into $2^{n(R_{11}-L_{11})}$ cells. For each
  given $w_{11}$, we can find the corresponding cell,then, we
  randomly choose a sub-bin from that cell, thus the sub-bin index
  $i^{'}$ can be decided. The codeword $v_{1}^{n}(i,i^{'}.i^{''})$
  will be chosen properly from that sub-bin.
  \item If $L_{11}+L_{12}\leq R_{11}\leq L_{11}+L_{12}+L_{3}$, then
  each sub-bin is mapped to at least one message $w_{11}$,
  therefore, given $w_{11}$, $i^{'}$ cab be decided. In each
  sub-bin, there are $2^{n(R_{11}-L_{11}-L_{12})}$ messages. The codeword $v_{1}^{n}(i,i^{'}.i^{''})$
  will be chosen randomly and properly from that sub-bin.
\end{enumerate}

Given $w_{22}$, we select $v_{2}^{n}(j,j^{'},j^{''})$ in the exact
same manner. From the given sub-bins , the encoder chooses the
codeword pair
$(v_{1}^{n}(i,i^{'},i^{''}),v_{2}^{n}(j,j^{'},j^{''}))$ that
satisfies the following property,

\begin{equation}\nonumber
(v_{1}^{n}(i,i^{'},i^{''}),v_{2}^{n}(j,j^{'},j^{''}))\in
A^{(n)}_{\epsilon}(V_{1},V_{2},U)
\end{equation}
where $A^{(n)}_{\epsilon}(U,V_{1},V_{2})$ denotes the set of jointly
typical sequences $u^{n}$, $v_{1}^{n}$, and $v_{2}^{n}$ with respect
to $P(u,v_{1},v_{2})$. If there is more than one such pair, the
transmitter randomly chooses one; if there is no such pair, an error
is declared.

Given $v_{1}^{n}$ and $v_{2}^{n}$, the channel input $x^{n}$ is
generated i.i.d. according to the distribution
$P(x^{n}|v_{1}^{n},v_{2}^{n})=\prod_{i=1}^{n}P(x_{i}|v_{1i},v_{2i})$.

3) \textit{Decoding}: The received signals at the legitimate
receivers, $y_{1}^{n}$ and $y_{2}^{n}$, are the outputs of the
channels $P(y_{1}^{n}|x^{n})=\prod_{i=1}^{n}P(y_{1,i}|x_{i})$ and
$P(y_{2}^{n}|x^{n})=\prod_{i=1}^{n}P(y_{2,i}|x_{i})$, respectively.
The first receiver looks for the unique sequence $u^{n}(k)$ such
that
\begin{equation}\nonumber
(u^{n}(k),y_{1}^{n})\in A^{(n)}_{\epsilon}(U,Y_{1}).
\end{equation}
If such $u^{n}(k)$ exists and is unique, set $\hat{k}=k$; otherwise,
declare an error. Upon decoding $k$, this receiver looks for
sequences $v_{1}^{n}(i,i^{'},i^{''})$ such that
\begin{equation}\nonumber
(v_{1}^{n}(i,i^{'},i^{''}),u^{n}(k),y_{1}^{n})\in
A^{(n)}_{\epsilon}(V_{1},U,Y_{1}).
\end{equation}
If such $v_{1}^{n}(i,i^{'},i^{''})$ exists and is unique, set
$\hat{i}=i$, $\hat{i}^{'}=i^{'}$, and $\hat{i}^{''}=i^{''}$;
otherwise, declare an error. Using the values of
$\hat{k},\hat{i},\hat{i}^{'}$ and $\hat{i}^{''}$, the decoder can
calculate the message indices $\hat{w_{0}},\hat{w}_{10}$ and
$\hat{w}_{11}$. The decoding for the second decoder is similar.

4) \textit{Error Probability Analysis}: Since the region of
$\mathbb{R}_{I}$ is a subset of the Marton's region, then the error
probability analysis is the same as \cite{33}.

5) \textit{Equivocation Calculation}: To meet the secrecy
requirements, we need to prove that the common message $W_{0}$, the
combination of $(W_{0},W_{1})$, the combination of $(W_{0},W_{2})$,
and the combination of $(W_{0},W_{1},W_{2})$  are perfectly secured.
The proof of secrecy requirement for the message $W_{0}$ is
straightforward and is therefore omitted.

To prove the secrecy requirement for $(W_{0},W_{1})$, we have
\begin{eqnarray}\nonumber
nR_{e10}&=&  H(W_{1},W_{0}|Z^{n})\\
\nonumber &=& H(W_{1},W_{0},Z^{n})-H(Z^{n})\\
\nonumber &=& H(W_{1},W_{0},U^{n},V_{1}^{n},Z^{n})-H(U^{n},V_{1}^{n}|W_{1},W_{0},Z^{n})-H(Z^{n})\\
\nonumber&=& H(W_{1},W_{0},U^{n},V_{1}^{n})+
H(Z^{n}|W_{1},W_{0},U^{n},V_{1}^{n})-H(U^{n}|W_{1},W_{0},Z^{n})-H(V_{1}^{n}|W_{1},W_{0},Z^{n},U^{n})-H(Z^{n})\\
\nonumber &\stackrel{(a)}{\geq}& H(W_{1},W_{0},U^{n},V_{1}^{n})+H(Z^{n}|W_{1},W_{0},U^{n},V_{1}^{n})-n\epsilon_{n}-H(Z^{n})\\
\nonumber &\stackrel{(b)}{=}&H(W_{1},W_{0},U^{n},V_{1}^{n})+H(Z^{n}|U^{n},V_{1}^{n})-n\epsilon_{n}-H(Z^{n})\\
\nonumber&\stackrel{(c)}{\geq}&
H(U^{n},V_{1}^{n})+H(Z^{n}|U^{n},V_{1}^{n})-n\epsilon_{n}-H(Z^{n})\\
\nonumber &=&
H(U^{n})+H(V_{1}^{n}|U^{n})-I(U^{n},V_{1}^{n};Z^{n})- n\epsilon_{n}\\
\nonumber &\stackrel{(d)}{\geq}&
\min\{I(U^{n};Y_{1}^{n}),I(U^{n};Y_{2}^{n})\}+I(V_{1}^{n};Y_{1}^{n}|U^{n})-I(V_{1}^{n};Z^{n}|U^{n})-I(U^{n};Z^{n})- n\epsilon_{n}\\
\nonumber &\stackrel{(e)}{\geq}& nR_{1}+nR_{0}-n\epsilon_{n},
\end{eqnarray}
where $(a)$ follows from Fano's inequality that bounds the term
$H(U^{n}|W_{1},W_{0},Z^{n})\leq
h(P_{we0}^{(n)})+nP_{we0}^{n}R_{w0}\leq n\epsilon_{n}/2$ and the
term $H(V_{1}^{n}|W_{1},W_{0},Z^{n},U^{n})\leq
h(P_{we1}^{(n)})+nP_{we1}^{n}R_{w1}\leq n\epsilon_{n}/2$ for
sufficiently large $n$. Here $P_{we0}^{n}$ and $P_{we1}^{n}$ denotes
the wiretapper's error probability of decoding $u^{n}$ and
$V_{1}^{n}$ in the case that the bin numbers $w_{0}$ and $w_{1}$ are
known to the eavesdropper, respectively. The eavesdropper first
looks for the unique $u^{n}$ in bin $w_{0}$ of the first layer, such
that it is jointly typical with $z^{n}$. As the number of candidate
codewords is small enough, the probability of error is arbitrarily
small for a sufficiently large $n$. Next, given $u^{n}$, the
eavesdropper looks for the unique $v_{1}^{n}$ in  the bin $w_{1}$
which is jointly typical with $z^{n}$. Similarly, since the number
of available candidates is small enough, then the probability of
decoding error is arbitrarily small. $(b)$ follows from the fact
that $(W_{1},W_{0})\rightarrow U^{n}\rightarrow V_{1}^{n}\rightarrow
Z^{n}$ forms a Markov chain. Therefore, we have
$I(W_{1},W_{0};Z^{n}|U^{n},V_{1}^{n})=0$, where it is implied that
$H(Z^{n}|W_{1},W_{0},U^{n},V_{1}^{n})=H(Z^{n}|U^{n},V_{1}^{n})$.
$(c)$ follows from the fact that $H(W_{1},W_{0},U^{n},X^{n})\geq
H(U^{n},X^{n})$. $(d)$ follows from that fact that $H(U^{n})\geq
\min\{I(U^{n};Y_{1}^{n}),I(U^{n};Y_{2}^{n})\}$ and
$H(V_{1}^{n}|U^{n})\geq I(V_{1}^{n};Y_{1}^{n}|U^{n})$. $(e)$ follows
from Lemma \ref{lem4} of the appendix \ref{app4}.

By using the same approach it is easy to show that,
\begin{eqnarray}\nonumber
nR_{e20}&=&H(W_{2},W_{0}|Z^{n})\\
\nonumber &\geq& nR_{2}+nR_{0}-n\epsilon_{n}.
\end{eqnarray}
Therefore, we only need to prove that $(W_{0},W_{1},W_{2})$ is
perfectly secured; we have
\begin{eqnarray}\nonumber \label{l4}
nR_{e120}&=&H(W_{1},W_{2},W_{0}|Z^{n})\\
\nonumber  &=& H(W_{1},W_{2},W_{0},Z^{n})-H(Z^{n})\\
\nonumber &=& H(W_{1},W_{2},W_{0},U^{n},V_{1}^{n},V_{2}^{n},Z^{n})- H(U^{n},V_{1}^{n},V_{2}^{n}|W_{1},W_{2},W_{0},,Z^{n})-H(Z^{n})\\
\nonumber &=&H(W_{1},W_{2},W_{0},U^{n},V_{1}^{n},V_{2}^{n})+ H(Z^{n}|W_{1},W_{2},W_{0},U^{n},V_{1}^{n},V_{2}^{n})- H(U^{n},V_{1}^{n},V_{2}^{n}|W_{1},W_{2},W_{0},Z^{n})\\ \nonumber &-&H(Z^{n})\\
\nonumber &\stackrel{(a)}{\geq}&
H(W_{1},W_{2},W_{0},U^{n},V_{1}^{n},V_{2}^{n})+H(Z^{n}|W_{1},W_{2},W_{0},U^{n},V_{1}^{n},V_{2}^{n})-n\epsilon_{n}-H(Z^{n})\\
\nonumber
&\stackrel{(b)}{=}&H(W_{1},W_{2},W_{0},U^{n},V_{1}^{n},V_{2}^{n})+H(Z^{n}|U^{n},V_{1}^{n},V_{2}^{n})-n\epsilon_{n}-H(Z^{n}) \\
\nonumber&\stackrel{(c)}{\geq}&
H(U^{n},V_{1}^{n},V_{2}^{n})+H(Z^{n}|U^{n},V_{1}^{n},V_{2}^{n})- n\epsilon_{n}-H(Z^{n}) \\
\nonumber &\stackrel{(d)}{=}& H(U^{n})+H(V_{1}^{n}|U^{n})+H(V_{2}^{n}|U^{n})-I(V_{1}^{n};V_{2}^{n}|U^{n})+H(Z^{n}|U^{n},V_{1}^{n},V_{2}^{n})-n\epsilon_{n}-H(Z^{n})\\
\nonumber &\stackrel{(e)}{\geq}&
\min\{I(U^{n};Y_{1}^{n}),I(U^{n};Y_{2}^{n})\}+I(V_{1}^{n};Y_{1}^{n}|U^{n})+I(V_{2}^{n};Y_{2}^{n}|U^{n})-I(V_{1}^{n};V_{2}^{n}|U^{n})-I(U^{n},V_{1}^{n},V_{2}^{n};Z^{n})\\ \nonumber &-&n\epsilon_{n}\\
\nonumber &\stackrel{(f)}{\geq}&
\min\{I(U^{n};Y_{1}^{n}),I(U^{n};Y_{2}^{n})\}+I(V_{1}^{n};Y_{1}^{n}|U^{n})+I(V_{2}^{n};Y_{2}^{n}|U^{n})-I(V_{1}^{n};V_{2}^{n}|U^{n})-I(V_{1}^{n},V_{2}^{n};Z^{n}|U^{n})\\ \nonumber &-&I(U^{n};Z^{n})-n\epsilon_{n}\\
\nonumber &\stackrel{(g)}{\geq}& n\min\{I(U;Y_{1}),I(U;Y_{2})\}+nI(V_{1};Y_{1}|U)+nI(V_{2};Y_{2}|U)-nI(V_{1};V_{2}|U)-nI(V_{1},V_{2};Z|U)\\ \nonumber &-& nI(U;Z)-n\epsilon_{n}\\
\nonumber &\geq& nR_{1}+nR_{2}+nR_{0}-n\epsilon_{n},
\end{eqnarray}
where $(a)$ follows from Fano's inequality, which states that for
sufficiently large $n$,
$H(U^{n},V_{1}^{n},V_{2}^{n}|W_{1},W_{2},W_{0},Z^{n})$ $\leq
h(P_{we}^{(n)})$ $+nP_{we}^{n}R_{w}\leq n\epsilon_{n}$. Here
$P_{we}^{n}$ denotes the wiretapper's error probability of decoding
$(u^{n},v_{1}^{n},v_{2}^{n})$ in the case that the bin numbers
$w_{0}$, $w_{1}$, and $w_{2}$ are known to the eavesdropper. Since
the sum rate is small enough, then $P_{we}^{n}\rightarrow 0$ for
sufficiently large $n$. $(b)$ follows from the following Markov
chain: $(W_{1},W_{2},W_{0})\rightarrow
(U^{n},V_{1}^{n},V_{2}^{n})\rightarrow$ $Z^{n}$. Hence, we have
$H(Z^{n}|W_{1},W_{2},W_{0},U^{n},V_{1}^{n},V_{2}^{n})=H(Z^{n}|U^{n},V_{1}^{n},V_{2}^{n})$.
$(c)$ follows from the fact that
$H(W_{1},W_{2},W_{0},U^{n},V_{1}^{n},V_{2}^{n})\geq
H(U^{n},V_{1}^{n},V_{2}^{n})$. $(d)$ follows from that fact that
$H(U^{n},V_{1}^{n},V_{2}^{n})=
H(U^{n})+H(V_{1}^{n}|U^{n})+H(V_{2}^{n}|U^{n})-I(V_{1}^{n};V_{2}^{n}|U^{n})$.
$(e)$ follows from the fact that $H(U^{n})\geq
\min\{I(U^{n};Y_{1}^{n}),I(U^{n};Y_{2}^{n})\}$ and
$H(V_{i}^{n}|U^{n})\geq I(V_{i}^{n};Y_{i}^{n}|U^{n})$ for $i=1,2$.
$(f)$ follows from the fact that
$I(U^{n},V_{1}^{n},V_{2}^{n};Z^{n})=I(U^{n};Z^{n})+I(V_{1}^{n},V_{2}^{n};Z^{n}|U^{n})$.
$(g)$ follows from Lemma \ref{lem3} and Lemma \ref{lem4} in the
appendix \ref{app4}. This completes the achievability proof.
\section{Proof of Theorem \ref{th2}}\label{app2}
\textit{Achievablity}: We need to show that the region of (\ref{l7})
and (\ref{l7_1}) is a subset of the achievability region of Theorem
\ref{th1}. In the achievability scheme of Theorem \ref{th1}, if we
set $\mathcal{W}_{2}=\emptyset$ and rename $W_{0}$ with $W_{2}$,
then using the degradedness, we obtain the following region,
\begin{IEEEeqnarray}{rl}
R_{1}+R_{2}&\leq I(V;Y_{1}|U)-I(V;Z|U)+I(U;Y_{2})-I(U;Z),\\
\nonumber R_{2}&\leq I(U;Y_{2})-I(U;Z).
\end{IEEEeqnarray}
Note that since the first receiver decodes both messages, the total
rate of this receiver is $R_{1}\leftarrow R_{1}+R_{2}$ and we have
\begin{IEEEeqnarray}{rl}
R_{1}&\leq I(UV;Y_{1}|U)+I(U;Y_{2})-I(UV;Z),\\
\nonumber R_{2}&\leq I(U;Y_{2})-I(U;Z).
\end{IEEEeqnarray}
Now, since $U\rightarrow V \rightarrow X \rightarrow
Y_{2}\rightarrow Z$ is a markov chain, then the following region is
a subset of the above region, and consequently, it is achievable,
\begin{IEEEeqnarray}{rl}
R_{1}&\leq I(X;Y_{1}|U)+I(U;Z)-I(X;Z),\\
\nonumber R_{2}&\leq I(U;Y_{2})-I(U;Z).
\end{IEEEeqnarray}
which is the same as that of region (\ref{l7}) and (\ref{l7_1}).
This completes the achievability proof.

 \textit{Converse}: The
transmitter sends two independent secret messages $W_{1}$ and
$W_{2}$ to Receiver $1$ and Receiver $2$, respectively. Let us
define $U_{i}=(W_{2},Y_{1}^{i-1})$. The following Lemma bounds the
secrecy rates for a general case of $(W_{1},W_{2})\rightarrow X^{n}
\rightarrow Y_{1}^{n}Y_{2}^{n}Z^{n}$:
\begin{lem} \label{ll1}
For the broadcast channel with an eavesdropper, the perfect secrecy
rates are bounded as follows,
\begin{IEEEeqnarray}{rl}\nonumber
    nR_{1}&\leq \sum_{i=1}^{n}I(W_{1};Y_{1i}|W_{2},Z_{i},Y_{1}^{i-1},\widetilde{Z}^{i+1})+n\delta_{1}+n\epsilon_{3},
    \\ \nonumber
    nR_{2}&\leq \sum_{i=1}^{n}I(W_{2};Y_{2i}|Z_{i},Y_{2}^{i-1},\widetilde{Z}^{i+1})+n\delta_{1}+n\epsilon_{2}.
\end{IEEEeqnarray}
\end{lem}
\begin{proof}
We need to prove the second bound. The first bound can similarly be
proven. $nR_{2}$ is bounded as follows:
\begin{eqnarray}\nonumber
nR_{2} &\stackrel{(a)}{\leq}& H(W_{2}|Z^{n})+n\epsilon_{2} \\
\nonumber &\stackrel{(b)}{\leq}&
H(W_{2}|Z^{n})-H(W_{2}|Y_{2}^{n})+n\delta_{1}+n\epsilon_{2}\\
\nonumber &=&
I(W_{2};Y_{2}^{n})-I(W_{2};Z^{n})+n\delta_{1}+n\epsilon_{2}
\end{eqnarray}
where $(a)$ follows from the secrecy constraint that
$H(W_{2}|Z^{n})\geq H(W_{2})-n\epsilon_{2}$. $(b)$ follows from
Fano's inequality that $H(W_{2}|Y_{2}^{n}) \leq n\delta_{1}$.
Next, we expand $I(W_{2};Y_{2}^{n})$ and $I(W_{2};Z^{n})$ as
follows.
\begin{eqnarray}\nonumber
I(W_{2};Y_{2}^{n}) &=&\sum_{i=1}^{n}I(W_{2};Y_{2i}|Y_{2}^{i-1}) \\
\nonumber
&=&\sum_{i=1}^{n}I(W_{2},\widetilde{Z}^{i+1};Y_{2i}|Y_{2}^{i-1})-I(\widetilde{Z}^{i+1};Y_{2i}|W_{2},Y_{2}^{i-1})\\
\nonumber &=&\sum_{i=1}^{n}
I(W_{2};Y_{2i}|Y_{2}^{i-1},\widetilde{Z}^{i+1})+I(\widetilde{Z}^{i+1};Y_{2i}|Y_{2}^{i-1})-I(\widetilde{Z}^{i+1};Y_{2i}|W_{2},Y_{2}^{i-1})\\
\nonumber
&=&\sum_{i=1}^{n}I(W_{2};Y_{2i}|Y_{2}^{i-1},\widetilde{Z}^{i+1})+\Delta_{1}-\Delta_{2},
\end{eqnarray}
where,
$\Delta_{1}=\sum_{i=1}^{n}I(\widetilde{Z}^{i+1};Y_{2i}|Y_{2}^{i-1})$
and
$\Delta_{2}=\sum_{i=1}^{n}I(\widetilde{Z}^{i+1};Y_{2i}|W_{2},Y_{2}^{i-1})$.
Similarly, we have,
\begin{eqnarray}\nonumber
I(W_{2};Z^{n}) &=&\sum_{i=1}^{n}I(W_{2};Z_{i}|\widetilde{Z}^{i+1}) \\
\nonumber
&=&\sum_{i=1}^{n}I(W_{2},Y_{2}^{i-1};Z_{i}|\widetilde{Z}^{i+1})-I(Y_{2}^{i-1};Z_{i}|W_{2},\widetilde{Z}^{i+1})\\
\nonumber &=&\sum_{i=1}^{n}
I(W_{2};Z_{i}|Y_{2}^{i-1},\widetilde{Z}^{i+1})+I(Y_{2}^{i-1};Z_{i}|\widetilde{Z}^{i+1})-I(Y_{2}^{i-1};Z_{i}|W_{2},\widetilde{Z}^{i+1})\\
\nonumber
&=&\sum_{i=1}^{n}I(W_{2};Z_{i}|Y_{2}^{i-1},\widetilde{Z}^{i+1})+\Delta_{1}^{*}-\Delta_{2}^{*},
\end{eqnarray}
where,
$\Delta_{1}^{*}=\sum_{i=1}^{n}I(Y_{2}^{i-1};Z_{i}|\widetilde{Z}^{i+1})$
and
$\Delta_{2}^{*}=\sum_{i=1}^{n}I(Y_{2}^{i-1};Z_{i}|W_{2},\widetilde{Z}^{i+1})$.
According to Lemma $7$ of \cite{4}, $\Delta_{1}=\Delta_{1}^{*}$ and
$\Delta_{2}=\Delta_{2}^{*}$. Thus, we have,
\begin{eqnarray}\nonumber
nR_{2}&\leq&
\sum_{i=1}^{n}I(W_{2};Y_{2i}|Y_{2}^{i-1},\widetilde{Z}^{i+1})-I(W_{2};Z_{i}|Y_{2}^{i-1},\widetilde{Z}^{i+1})+n\delta_{1}+n\epsilon_{2}
\\ \nonumber &=&\sum_{i=1}^{n}H(W_{2}|Z_{i},Y_{2}^{i-1},\widetilde{Z}^{i+1})-H(W_{2}|Y_{2i},Y_{2}^{i-1},\widetilde{Z}^{i+1})+n\delta_{1}+n\epsilon_{2}\\
\nonumber
&\stackrel{(a)}{\leq}&\sum_{i=1}^{n}H(W_{2}|Z_{i},Y_{2}^{i-1},\widetilde{Z}^{i+1})-H(W_{2}|Y_{2i},Z_{i},Y_{2}^{i-1},\widetilde{Z}^{i+1})+n\delta_{1}+n\epsilon_{2}\\
\nonumber
&=&\sum_{i=1}^{n}I(W_{2};Y_{2i}|Z_{i},Y_{2}^{i-1},\widetilde{Z}^{i+1})+n\delta_{1}\nonumber+n\epsilon_{2},
\end{eqnarray}
where $(a)$ follows from the fact that conditioning always decreases
the entropy.
\end{proof}
Now according to the above Lemma, the secrecy rates are bounded as
follows:
\begin{eqnarray}\nonumber
nR_{1} &\stackrel{(a)}{\leq}&
\sum_{i=1}^{n}I(W_{1};Y_{1,i}|W_{2},Z_{i},Y_{1}^{i-1},\widetilde{Z}^{i+1})+ n\delta_{1}+n\epsilon_{3}\\
\nonumber &=&
 \sum_{i=1}^{n}I(W_{1};Y_{1,i}|U_{i},Z_{i},\widetilde{Z}^{i+1})+n\delta_{1}+n\epsilon_{3}\\
\nonumber &\stackrel{(b)}{\leq}& \sum_{i=1}^{n} I(X_{i};Y_{1,i}|U_{i},Z_{i},\widetilde{Z}^{i+1})+n\delta_{1}+n\epsilon_{3}\\
\nonumber &\stackrel{(c)}{=}&
\sum_{i=1}^{n}I(X_{i};Y_{1,i},U_{i},Z_{i}|\widetilde{Z}^{i+1})-I(X_{i};Z_{i}|\widetilde{Z}^{i+1})-I(X_{i};U_{i}|Z_{i},\widetilde{Z}^{i+1}) +n\delta_{1}+n\epsilon_{3}\\
\nonumber&\stackrel{(d)}{=}& \sum_{i=1}^{n}
I(X_{i};Y_{1,i}|U_{i},\widetilde{Z}^{i+1})+I(X_{i};U_{i}|\widetilde{Z}^{i+1})-I(X_{i};Z_{i}|\widetilde{Z}^{i+1})-I(X_{i};U_{i}|Z_{i},\widetilde{Z}^{i+1})+n\delta_{1}+n\epsilon_{3}\\
\nonumber &\stackrel{(e)}{=}& \sum_{i=1}^{n}
I(X_{i};Y_{1,i}|U_{i},\widetilde{Z}^{i+1})-I(X_{i};Z_{i}|\widetilde{Z}^{i+1})+I(Z_{i};U_{i}|\widetilde{Z}^{i+1})-I(Z_{i};U_{i}|X_{i},\widetilde{Z}^{i+1})
+ n\delta_{1}+n\epsilon_{3}\\
\nonumber &\stackrel{(f)}{=}&
\sum_{i=1}^{n}I(X_{i};Y_{1,i}|U_{i},\widetilde{Z}^{i+1})-I(X_{i};Z_{i}|\widetilde{Z}^{i+1})+I(Z_{i};U_{i}|\widetilde{Z}^{i+1})+n\delta_{1}+n\epsilon_{3},
\end{eqnarray}
where $(a)$ follows from the Lemma (\ref{ll1}). $(b)$ follows from
the data processing theorem. $(c)$ follows from the chain rule.
$(d)$ follows from the fact that
$I(X_{i};Y_{1,i},U_{i},Z_{i}|\widetilde{Z}^{i+1})=I(X_{i};U_{i}|\widetilde{Z}^{i+1})
+I(X_{i};Y_{1,i}|U_{i},\widetilde{Z}^{i+1})+I(X_{i};Z_{i}|Y_{1,i},U_{i},\widetilde{Z}^{i+1})$
and from the fact that $\widetilde{Z}^{i+1}U_{i}\rightarrow X_{i}
\rightarrow Y_{1,i} \rightarrow Y_{2,i} \rightarrow Z_{i}$ forms a
Markov chain, which means that
$I(X_{i};Z_{i}|Y_{1,i},U_{i},\widetilde{Z}^{i+1})=0$. $(e)$ follows
from the fact that
$I(X_{i};U_{i}|\widetilde{Z}^{i+1})-I(X_{i};U_{i}|Z_{i},\widetilde{Z}^{i+1})=I(Z_{i};U_{i}|\widetilde{Z}^{i+1})-I(Z_{i};U_{i}|X_{i},\widetilde{Z}^{i+1})$.
$(f)$ follows from the fact that $\widetilde{Z}^{i+1}
U_{i}\rightarrow X_{i} \rightarrow Z_{i}$ forms a Markov chain.
Thus, $I(Z_{i};U_{i}\widetilde{Z}^{i+1}|X_{i})=0$ which implies that
$I(Z_{i};U_{i}|X_{i},\widetilde{Z}^{i+1})=0$.

For the second receiver, we have
\begin{eqnarray}\nonumber
nR_{2} &\stackrel{(a)}{\leq}&\sum_{i=1}^{n}I(W_{2};Y_{2,i}|Y_{2}^{i-1},Z_{i},\widetilde{Z}^{i+1})+n\delta_{2}+n\epsilon_{1}\\
\nonumber
&=&\sum_{i=1}^{n}H(Y_{2,i}|Y_{2}^{i-1},Z_{i},\widetilde{Z}^{i+1})-H(Y_{2,i}|W_{2},Y_{2}^{i-1},Z_{i},\widetilde{Z}^{i+1})+n\delta_{2}+n\epsilon_{1}\\
\nonumber &\stackrel{(b)}{\leq}&\sum_{i=1}^{n}H(Y_{2,i}|Z_{i},\widetilde{Z}^{i+1})-H(Y_{2,i}|W_{2},Y_{1}^{i-1},Y_{2}^{i-1},Z_{i},\widetilde{Z}^{i+1})+n\delta_{2}+n\epsilon_{1}\\
\nonumber &\stackrel{(c)}{=}&\sum_{i=1}^{n}H(Y_{2,i}|Z_{i},\widetilde{Z}^{i+1})-H(Y_{2,i}|U_{i},Z_{i},\widetilde{Z}^{i+1})+n\delta_{2}+n\epsilon_{1}\\
\nonumber &=&\sum_{i=1}^{n}I(Y_{2,i};U_{i}|Z_{i},\widetilde{Z}^{i+1})+n\delta_{2}+n\epsilon_{1}\\
\nonumber &=&\sum_{i=1}^{n}I(Y_{2,i};U_{i}|\widetilde{Z}^{i+1})+I(Y_{2,i};Z_{i}|U_{i},\widetilde{Z}^{i+1})-I(Y_{2,i};Z_{i}|\widetilde{Z}^{i+1})+n\delta_{2}+n\epsilon_{1}\\
\nonumber &=&\sum_{i=1}^{n}I(Y_{2,i};U_{i}|\widetilde{Z}^{i+1})-I(Z_{i};U_{i}|\widetilde{Z}^{i+1})+I(Z_{i};U_{i}|Y_{2,i},\widetilde{Z}^{i+1})+n\delta_{2}+n\epsilon_{1}\\
\nonumber
&\stackrel{(d)}{=}&\sum_{i=1}^{n}I(Y_{2,i};U_{i}|\widetilde{Z}^{i+1})-I(Z_{i};U_{i}|\widetilde{Z}^{i+1})+n\delta_{2}+n\epsilon_{1}\nonumber,
\end{eqnarray}
where $(a)$ follows from the lemma (\ref{ll1}). $(b)$ follows from
the fact that conditioning always decreases the entropy. $(c)$
follows from the fact that $Y_{2}^{i-1}\rightarrow
W_{2}\widetilde{Z}^{i+1}Y_{1}^{i-1}\rightarrow Y_{2i}\rightarrow
Z_{i}$ forms a Markov chain. $(d)$ follows from the fact that
$\widetilde{Z}^{i+1}U_{i}\rightarrow Y_{2,i} \rightarrow Z_{i}$
forms a Markov chain. Thus
$I(Z_{i};U_{i}\widetilde{Z}^{i+1}|Y_{2i})=0$ which implies that
$I(Z_{i};U_{i}|Y_{2i},\widetilde{Z}^{i+1})=0$. Now, following
\cite{34}, let us define the time sharing random variable $Q$ which
is uniformly distributed over $\{1,2,...,n\}$ and independent of
$(W_{1},W_{2},X^{n},Y_{1}^{n},Y_{2}^{n})$. Let us define $U=U_{Q},~
V=(\widetilde{Z}^{Q+1},Q),~ X=X_{Q},~ Y_{1}=Y_{1,Q},~
Y_{2}=Y_{2,Q},~ Z=Z_{Q}$, then $R_{1}$ and $R_{2}$ can be written as
\begin{IEEEeqnarray}{rl}
    R_{1}&\leq I(X;Y_{1}|U,V)+I(U;Z|V)-I(X;Z|V), \\
    R_{2}&\leq I(U;Y_{2}|V)-I(U;Z|V).
\end{IEEEeqnarray}
Note that the boundary of this region is characterized by the
maximization of $R_{1}+\mu R_{2}$ over this region for $\mu \geq 1$.
On the other hand we have,
\begin{equation}
R_{1}+\mu R_{2}\leq I(X;Y_{1}|U,V)+I(U;Z|V)-I(X;Z|V)+\mu\left(
I(U;Y_{2}|V)-I(U;Z|V)\right)
\end{equation}
Since conditional mutual information is the average of the
unconditional ones, the largest region is achieved when $V$ is a
constant. This proves the converse part.
\section{Proof of Theorem \ref{th3}}\label{app3}

\textit{Achievability}: Let $U\sim \mathcal{N}(0,(1-\alpha)P)$ and
$X^{'}\sim \mathcal{N}(0,\alpha P)$ be independent and $X=U+X^{'}
\sim \mathcal{N}(0,P)$. Now consider the following secure
superposition coding scheme:

1) \textit{Codebook Generation}: Generate $2^{nI(U;Y_{2})}$ i.i.d
Gaussian codewords $u^{n}$ with average power $(1-\alpha)P$ and
randomly distribute these codewords into $2^{nR_{2}}$ bins. Then
index each bin by $w_{2}\in\{1,2,...,2^{nR_{2}}\}$. Generate an
independent set of $2^{nI(X^{'};Y_{1}})$ i.i.d Gaussian codewords
$x^{'n}$ with average power $\alpha P$. Then, randomly distribute
them into $2^{nR_{1}}$ bins. Index each bin by
$w_{1}\in\{1,2,...,2^{nR_{1}}\}$.

2) \textit{Encoding}: To send messages $w_{1}$ and $w_{2}$, the
transmitter randomly chooses one of the codewords in bin $w_{2}$,
(say $u^{n}$) and one of the codewords in bin $w_{1}$ (say $x^{'n}$
). The transmitter then simply transmits $x^{n} =u^{n}+ x^{'n}$.

3) \textit{Decoding}: The received signal at the legitimate
receivers are $y_{1}^{n}$ and $y_{2}^{n}$, respectively. Receiver
$2$ determines the unique $u^{n}$ such that $(u^{n},y_{2}^{n})$ are
jointly typical and declares the index of the bin containing $u^{n}$
as the message received. If there is none of such or more than one
of such, an error is declared. Receiver $1$ uses the successive
cancelation method; it first decodes $u^{n}$ and subtracts it from
$y_{1}^{n}$ and then looks for the unique $x^{'n}$ such that
$(x^{'n},y_{1}^{n}-u^{n})$ are jointly typical and declares the
index of the bin containing $x^{'n}$ as the message received.

It can be shown that if $R_{1}$ and $R_{2}$ satisfy (\ref{g1}) and
(\ref{g2}), the error probability analysis and equivocation
calculation is straightforward and may therefore be omitted.

\textit{Converse}: According to the previous section, $R_{2}$ is
bounded as follows:

\begin{eqnarray}\label{l10}
nR_{2} \leq I(Y_{2}^{n};U^{n}|Z^{n}) =
h(Y_{2}^{n}|Z^{n})-h(Y_{2}^{n}|U^{n},Z^{n}),
\end{eqnarray}
where $h$ is the differential entropy. The classical entropy power
inequality states that:
\begin{eqnarray}\nonumber
2^{\frac{2}{n}h(Y_{2}^{n}+N_{3}^{'n})}\geq
2^{\frac{2}{n}h(Y_{2}^{n})}+2^{\frac{2}{n}h(N_{3}^{'n})}
\end{eqnarray}
Therefore, $h(Y_{2}^{n}|Z^{n})$ may be written as follows:
\begin{eqnarray}\nonumber
h(Y_{2}^{n}|Z^{n})&=&h(Z^{n}|Y_{2}^{n})+h(Y_{2}^{n})-h(Z^{n})\\
\nonumber
&=& \frac{n}{2}\log2\pi e(\sigma_{3}^{2}-\sigma_{2}^{2})+h(Y_{2}^{n})-h(Y_{2}^{n}+N_{3}^{'n})\\
\nonumber &\leq&\frac{n}{2}\log2\pi
e(\sigma_{3}^{2}-\sigma_{2}^{2})+h(Y_{2}^{n})-\frac{n}{2}\log(2^{\frac{2}{n}h(Y_{2}^{n})}+2\pi
e(\sigma_{3}^{2}-\sigma_{2}^{2})).
\end{eqnarray}
On the other hand, for any fixed $a\in \mathcal{R}$, the function
\begin{eqnarray}\nonumber
f(t,a)=t-\frac{n}{2}\log(2^{\frac{2}{n}t}+a)
\end{eqnarray}
is concave in $t$ and has a global maximum at the maximum value of
$t$. Thus, $h(Y_{2}^{n}|Z^{n})$ is maximized when $Y_{2}^{n}$ (or
equivalently $X^{n}$) has Gaussian distribution. Hence,
\begin{eqnarray} \label{l8}\nonumber
h(Y_{2}^{n}|Z^{n})&\leq& \frac{n}{2}\log2\pi
e(\sigma_{3}^{2}-\sigma_{2}^{2})+\frac{n}{2}\log2\pi
e(P+\sigma_{2}^{2})-\frac{n}{2}\log2\pi e(P+\sigma_{3}^{2})\\
 &=&\frac{n}{2}\log\left(\frac{2\pi
e(\sigma_{3}^{2}-\sigma_{2}^{2})(P+\sigma_{2}^{2})}{P+\sigma^{2}_{3}}\right).
\end{eqnarray}
Note that another method to obtain (\ref{l8}) is using the worst
additive noise lemma (see \cite{39,40} for details). Now consider
the term $h(Y_{2}^{n}|U^{n},Z^{n})$. This term is lower bounded with
$h(Y_{2}^{n}|U^{n},X^{n},Z^{n})=\frac{n}{2}\log2\pi
e(\sigma_{2}^{2})$ which is greater than $\frac{n}{2}\log2\pi
e(\frac{\sigma_{2}^{2}(\sigma_{3}^{2}-\sigma_{2}^{2})}{\sigma_{3}^{2}})$.
Hence,
\begin{eqnarray}\label{l9}
\frac{n}{2}\log2\pi
e(\frac{\sigma_{2}^{2}(\sigma_{3}^{2}-\sigma_{2}^{2})}{\sigma_{3}^{2}})\leq
h(Y_{2}^{n}|U^{n},Z^{n})\leq h(Y_{2}^{n}|Z^{n}).
\end{eqnarray}
Inequalities (\ref{l8}) and (\ref{l9}) imply that there exists an
$\alpha \in [0,1]$ such that
\begin{eqnarray}\label{l11}
h(Y_{2}^{n}|U^{n},Z^{n})=\frac{n}{2}\log\left(\frac{2\pi
e(\sigma_{3}^{2}-\sigma_{2}^{2})(\alpha P+\sigma_{2}^{2})}{\alpha
P+\sigma_{3}^{2}}\right).
\end{eqnarray}
Substituting (\ref{l11}) and (\ref{l8}) into (\ref{l10}) yields the
desired bound
\begin{eqnarray}\nonumber
nR_{2} &\leq&  h(Y_{2}^{n}|Z^{n})-h(Y_{2}^{n}|U^{n},Z^{n}) \\
\nonumber &\leq&\frac{n}{2}\log\left(\frac{(P+\sigma_{2}^{2})(\alpha
P + \sigma_{3}^{2})}{(P+\sigma_{3}^{2})(\alpha P + \sigma_{2}^{2})}\right)\\
 &=& nC\left(\frac{(1-\alpha) P}{\alpha P
+\sigma_{2}^{2}}\right)-nC\left(\frac{(1-\alpha) P}{\alpha P
+\sigma_{3}^{2}}\right).
\end{eqnarray}
Note that the left hand side of (\ref{l11}) can be written as
$h(Y_{2}^{n},Z^{n}|U^{n})-h(Z^{n}|U^{n})$ which implies that
\begin{eqnarray}
h(Y_{2}^{n}|U^{n})-h(Z^{n}|U^{n})=\frac{n}{2}\log\left(\frac{\alpha
P+\sigma_{2}^{2}}{\alpha P+\sigma_{3}^{2}}\right).
\end{eqnarray}
Since $\sigma_{1}^{2}\leq \sigma_{2}^{2}\leq \sigma_{3}^{2}$, there
exists a $0\leq \beta \leq 1$ such that $\sigma_{2}^{2}=(1-\beta)
\sigma_{1}^{2}+\beta\sigma_{3}^{2}$, or equivalently,
$\sigma_{2}^{2}=
\sigma_{1}^{2}+\beta(\sigma_{3}^{2}-\sigma_{1}^{2})$. Therefore,
since $Y_{1}^{n}\rightarrow Y_{2}^{n} \rightarrow Z^{n}$ forms a
Markov chain, the received signals $Z^{n}$ and $Y_{2}^{n}$ can be
written as $Z^{n} =Y_{1}^{n} + \widetilde{N}^{n}$ and $Y_{2}^{n} =
Y_{1}^{n} + \sqrt{\beta} \widetilde{N}^{n}$ where $\widetilde{N}$ is
an independent Gaussian noise with variance $\widetilde{\sigma}^{2}
= \sigma_{3}^{2}-\sigma_{1}^{2}$.  All noises are Gaussian random
$n$-vector with a positive definite covariance matrix. Costa's
entropy power inequality \cite{41} states that (see also \cite{42}
for its linear version),
\begin{IEEEeqnarray}{rl}
2^{\frac{2}{n}h(Y_{1}^{n} + \sqrt{\beta}
\widetilde{N}^{n}|U^{n})}\geq
(1-\beta)2^{\frac{2}{n}h(Y_{1}^{n}|U^{n})}+\beta
2^{\frac{2}{n}h(Y_{1}^{n} + \widetilde{N}^{n}|U^{n})}
\end{IEEEeqnarray}
for any random $n$-vector $Y_{1}^{n}$ and Gaussian random $n$-vector
of $\widetilde{N}^{n}$. Equivalently we have,
\begin{IEEEeqnarray}{rl}\label{rr}
2^{\frac{2}{n}h(Y_{2}^{n}|U^{n})}\geq
(1-\beta)2^{\frac{2}{n}h(Y_{1}^{n}|U^{n})}+\beta
2^{\frac{2}{n}h(Z^{n}|U^{n})}
\end{IEEEeqnarray}
After some manipulations of (\ref{rr}), we obtain
\begin{IEEEeqnarray}{rl}\label{op} \nonumber
&h(Y_{1}^{n}|U^{n})-h(Z^{n}|U^{n}) \\
\nonumber & \leq
 \frac{n}{2}\log\left(\frac{\alpha
P+\sigma_{2}^{2}-\beta(\alpha P +\sigma_{3}^{2})}{(1-\beta)(\alpha P+\sigma_{3}^{2})}\right)\\
 &=\frac{n}{2}\log\left(\frac{\alpha P+\sigma_{1}^{2}}{\alpha
P+\sigma_{3}^{2}}\right).
\end{IEEEeqnarray}
 The rate $R_{1}$ is bounded as follows
\begin{eqnarray}
nR_{1}&\leq& I(X^{n};Y_{1}^{n}|U^{n})-I(X^{n};Z^{n})+I(U^{n};Z^{n})\\
\nonumber &=&h(Y_{1}^{n}|U^{n})-h(Y_{1}^{n}|X^{n},U^{n})+h(Z^{n}|X^{n})-h(Z^{n}|U^{n}) \\
\nonumber
&=&h(Y_{1}^{n}|U^{n})-h(Z^{n}|U^{n})+\frac{n}{2}\log(\frac{\sigma_{3}^{2}}{\sigma_{1}^{2}})\\
\nonumber &\stackrel{(a)}{\leq}&  \frac{n}{2}\log\left( \frac{\alpha
P + \sigma_{1}^{2}}{\alpha P + \sigma_{3}^{2}}\frac{\sigma_{3}^{2}}{\sigma_{1}^{2}} \right)\\
\nonumber &=& nC\left(\frac{\alpha
P}{\sigma_{1}^{2}}\right)-nC\left(\frac{\alpha
P}{\sigma_{3}^{2}}\right),
\end{eqnarray}
where $(a)$ follows from (\ref{op}).

\section{Complementary Lemmas for Equivocation Analysis}\label{app4}
\begin{lem}\label{lem3}
Assume $U^{n},V_{1}^{n},V_{2}^{n}$ and $Z^{n}$ are generated
according to the achievablity scheme of Theorem \ref{th1}, then we
have,
\begin{IEEEeqnarray}{lr}\nonumber
I(V_{1}^{n},V_{2}^{n};Z^{n}|U^{n})\leq nI(V_{1},V_{2};Z|U)+n\delta_{1n},\\
\nonumber I(V_{1}^{n};V_{2}^{n}|U^{n})\leq
nI(V_{1};V_{2}|U)+n\delta_{2n}.
\end{IEEEeqnarray}
\end{lem}
\begin{proof}
Let $A_{\epsilon}^{n}(P_{U,V_{1},V_{2},Z})$ denote the set of
typical sequences $(U^{n},V_{1}^{n},V_{2}^{n},Z^{n})$ with respect
to $P_{U,V_{1},V_{2},Z}$, and
\begin{equation}\nonumber
\zeta=\left\{
       \begin{array}{ll}
         1, & (U^{n},V_{1}^{n},V_{2}^{n},Z^{n})\notin A_{\epsilon}^{n}(P_{U,V_{1},V_{2},Z}); \\
         0, & \hbox{otherwise},
       \end{array}
     \right.
\end{equation}
be the corresponding indicator function. We expand
$I(V_{1}^{n},V_{2}^{n};Z^{n}|U^{n})$ as follow,
\begin{IEEEeqnarray}{rl}\label{l31}
I(V_{1}^{n},V_{2}^{n};Z^{n}|U^{n})&\leq
I(V_{1}^{n},V_{2}^{n},\zeta;Z^{n}|U^{n})\\ \nonumber
&=I(V_{1}^{n},V_{2}^{n};Z^{n}|U^{n},\zeta)+I(\zeta;Z^{n}|U^{n})\\
\nonumber
&=\sum_{j=0}^{1}P(\zeta=j)I(V_{1}^{n},V_{2}^{n};Z^{n}|U^{n},\zeta=j)+I(\zeta;Z^{n}|U^{n}).
\end{IEEEeqnarray}
According to the joint typicality property, we have
\begin{IEEEeqnarray}{rl}\label{l32}
P(\zeta=1)I(V_{1}^{n},V_{2}^{n};Z^{n}|U^{n},\zeta=1)&\leq
nP((U^{n},V_{1}^{n},V_{2}^{n},Z^{n})\notin
A_{\epsilon}^{n}(P_{U,V_{1},V_{2},Z}))\log\|\mathcal{Z}\|\\
\nonumber &\leq n\epsilon_{n}\log\|\mathcal{Z}\|.
\end{IEEEeqnarray}
Note that,
\begin{equation}\label{l33}
I(\zeta;Z^{n}|U^{n})\leq H(\zeta)\leq 1
\end{equation}
Now consider the term
$P(\zeta=0)I(V_{1}^{n},V_{2}^{n};Z^{n}|U^{n},\zeta=0)$. Following
the sequence joint typicality properties, we have
\begin{IEEEeqnarray}{rl}\label{l34}
P(\zeta=0)I(V_{1}^{n},V_{2}^{n};Z^{n}|U^{n},\zeta=0)&\leq
I(V_{1}^{n},V_{2}^{n};Z^{n}|U^{n},\zeta=0)\\ \nonumber &=
\sum_{(U^{n},V_{1}^{n},V_{2}^{n},Z^{n})\in
A_{\epsilon}^{n}}P(U^{n},V_{1}^{n},V_{2}^{n},Z^{n})\big(\log
P(V_{1}^{n},V_{2}^{n},Z^{n}|U^{n})-\log
P(V_{1}^{n},V_{2}^{n}|U^{n})\\ \nonumber &-\log P(Z^{n}|U^{n})\big),\\
\nonumber &\leq
n\left[-H(V_{1},V_{2},Z|U)+H(V_{1},V_{2}|U)+H(Z|U)+3\epsilon_{n}\right],\\
\nonumber &=n\left[I(V_{1},V_{2};Z|U)+3\epsilon_{n}\right].
\end{IEEEeqnarray}
By substituting (\ref{l32}), (\ref{l33}), and (\ref{l34}) into
(\ref{l31}), we get the desired reasult,
\begin{IEEEeqnarray}{rl}
I(V_{1}^{n},V_{2}^{n};Z^{n}|U^{n})&\leq
nI(V_{1},V_{2};Z|U)+n\left(\epsilon_{n}\log\|\mathcal{Z}\|+3\epsilon_{n}+\frac{1}{n}\right),\\
\nonumber &=nI(V_{1},V_{2};Z|U)+n\delta_{1n},
\end{IEEEeqnarray}
where,
\begin{equation}\nonumber
\delta_{1n}=\epsilon_{n}\log\|\mathcal{Z}\|+3\epsilon_{n}+\frac{1}{n}.
\end{equation}
Following the same steps, one can prove that
\begin{IEEEeqnarray}{lr}
I(V_{1}^{n};V_{2}^{n}|U^{n})\leq nI(V_{1};V_{2}|U)+n\delta_{2n}.
\end{IEEEeqnarray}
\end{proof}
Using the same approach as in Lemma \ref{lem3}, we can prove the
following lemmas.
\begin{lem}\label{lem4}
Assume $U^{n}, V_{1}^{n}, Y_{1}^{n}$ and $Y_{2}^{n}$ are generated
according to the achievablity scheme of Theorem \ref{th1}, then we
have,
\begin{IEEEeqnarray}{lr}\nonumber
I(V_{1}^{n};Y_{1}^{n}|U^{n})\leq nI(V_{1};Y_{1}|U)+n\delta_{3n},\\
\nonumber I(V_{1}^{n};Z^{n}|U^{n})\leq nI(V_{1};Z|U)+n\delta_{4n},\\
\nonumber I(U^{n};Z^{n})\leq nI(U;Z)+n\delta_{5n},\\ \nonumber
I(U^{n};Y_{1}^{n})\leq nI(U;Y_{1})+n\delta_{6n}\\ \nonumber
I(U^{n};Y_{2}^{n})\leq nI(U;Y_{2})+n\delta_{7n}
\end{IEEEeqnarray}
\end{lem}
\begin{proof}
The steps of the proof are very similar to the steps of proof of
Lemma \ref{lem3} and may be omitted here.
\end{proof}

\end{document}